# Filtering Microarray Correlations by Statistical Literature Analysis Yields Potential Hypotheses for Lactation Research

**Maurice HT Ling[1,2]** (mauriceling@acm.org)
**Christophe Lefevre [1,3,4]** (Chris.Lefevre@med.monash.edu.au)
**Kevin R Nicholas[1,4]** (kevin.nicholas@deakin.edu.au)

[1]CRC for Innovative Dairy Products, Department of Zoology, The University of Melbourne, Australia
[2]School of Chemical and Life Sciences, Singapore Polytechnic, Singapore
[3]Victorian Bioinformatics Consortium, Monash University, Australia
[4]Institute of Technology Research and Innovation, Deakin University, Australia

## Abstract

**Background**

Recent studies have demonstrated that the cyclical nature of mouse lactation[1] can be mirrored at the transcriptome[2] level of the mammary glands but making sense of microarray[3] results requires analysis of large amounts of biological information which is increasingly difficult to access as the amount of literature increases. Extraction of protein-protein interaction from text by statistical and natural language processing has shown to be useful in managing the literature. Correlations between gene expression across a series of samples is a simple method to analyze microarray data as it was found that genes that are related in functions exhibit similar expression profiles[4]. Microarrays had been used to examine the transcriptome of mouse lactation and found that the cyclic nature of the lactation cycle as observed histologically is reflected at the transcription level. However, there has been no study to date using text mining to sieve microarray analysis to generate new hypotheses for further research in the field of lactational biology.

**Results**

Our results demonstrated that a previously reported protein name co-occurrence method (5-mention PubGene) which was not based on a hypothesis testing framework, is generally more stringent than the 99$^{th}$ percentile of Poisson distribution-based method of calculating co-occurrence. It agrees with previous methods using natural language processing to extract protein-protein interaction from text as more than 96% of the interactions found by natural language processing methods to coincide with the results from 5-mention PubGene method. However, less than 2% of the gene co-expressions analyzed by microarray were found from direct co-occurrence or interaction information extraction from the literature. At the same time, combining microarray and literature analyses, we derive a novel set of 7 potential functional protein-protein interactions that had not been previously described in the literature.

---

1  Lactation is the process of milk production.

2  Transcriptome is the set of genes that are active in a given cell at any one time.

3  Microarray is a multiplex technology used in molecular biology to measure the activity of a set of genes at any one time.

4  A gene expression profile is the trend of activity for all the genes across different time points or conditions.

**Conclusions**

We conclude that the 5-mention PubGene method is more stringent than the 99$^{th}$ percentile of Poisson distribution method for extracting protein-protein interactions by co-occurrence of entity names and literature analysis may be a potential filter for microarray analysis to isolate potentially novel hypotheses for further research.

# 1. Background

Microarray technology is a transcriptome analysis tool which had been used in the study of the mouse lactation cycle (Clarkson and Watson, 2003; Rudolph et al., 2007). A number of advances in microarray analysis have been made recently. For example, inferring the underlying genetic network from microarray results (Rawool and Venkatesh, 2007; Maraziotis et al., 2007) by statistical correlation of gene expression across a series of samples (Reverter et al., 2005), then deriving functional network clusters by mapping onto Gene Ontology (Beissbarth, 2006). It has been shown that functionally related genes demonstrate similar expression profiles (Reverter et al., 2005). These methods have been used to study functional gene sets for basal cell carcinoma (O'Driscoll et al., 2006). The amount of information in published form is increasing exponentially, making it difficult for researchers to keep abreast with the relevant literature (Hunter and Cohen, 2006). At the same time, there has been no study to demonstrate that the current status of knowledge in protein-protein interactions in the literature is useful to increase the understanding of microarray data.

The two major streams for biomedical protein-protein information extraction are natural language processing (NLP) and co-occurrence statistics (Cohen and Hersh, 2005; Jensen et al., 2006). The main reason for concurrent existence of these two methods is their complementary effect in terms of information extraction (Jensen et al., 2006). NLP has a lower recall or sensitivity than co-occurrence but tends to be more precise compared with co-occurrence statistical methods (Wren and Garner, 2004; Jensen et al., 2006). Mathematically, precision is the number of true positives divided by the total number of items labeled by the system as positive (number of true positives divided by the sum of true and false positives), whereas recall is the number of true positives identified by the system divided the number of actual positives (number of true positives divided by the sum of true positives and false negatives). A number of tools have approached protein-protein interaction extraction from the NLP perspective, these include GENIES (Friedman et al., 2001), MedScan (Novichkova et al., 2003), PreBIND (Donaldson et al., 2003), BioRAT (David et al., 2004), GIS (Chiang et al., 2004), CONAN (Malik et al., 2006), and Muscorian (Ling et al., 2007). Muscorian (Ling et al., 2007) achieved at least 82% precision and 30% in recall (sensitivity). NLP methods made use of the grammatical forms of words and structure of a valid sentence to identify the grammatical roles of each word in a sentence, parse the sentence into phrases and extracting information such as subject-verb-object structures from these phrases. Co-occurrence, a statistical method, is based on the thesis that multiple occurrences of the same pair of entities suggests that the pair of entities are related in some way and the likelihood of such relatedness increases with higher co-occurrence. In another words, co-occurrence methods tend to view the text as a bag of un-sequenced words. Hence, depending on the threshold allowed, which will translate to the precision of the entire system, recall could be total, as implied in PubGene (Jenssen et al., 2001).

PubGene (Jenssen et al., 2001) defined interactions by co-occurrence to the simplest and widest possible form by assigning an interaction between 2 proteins if these 2 proteins appear in the same article just once in the entire library of 10 million articles and found that this criterion has 60% precision (1-Mention PubGene method). Although it was not stated in the article (Jenssen et al., 2001), it is obvious that such a criterion would yield 100% recall or sensitivity, giving an F-score of 0.75. F-score is defined as the harmonic mean of precision and recall, attributing equal weight to both precision and recall. However, 60% precision is usually unsatisfactory for most applications. PubGene (Jenssen et al., 2001) had also defined a "5-Mention" method which requires 5 or more articles with 2 protein names to assign an interaction with 72% precision. It is generally accepted that precision and recall are inversely related; hence, it can be expected that the "5-Mention" method will not be 100% sensitive. However, PubGene was benchmarked against the Database of Interacting Proteins and OMIM, making it more difficult to appreciate the statistical basis of "1-Mention" and "5-Mention" methods as compared to using a hypothesis testing framework in Chen et al. (2008). In addition, PubGene is unable to extract the nature of interactions, for example, binding or inhibiting interactions. On the other hand, NLP is designed to extract the nature of interactions (Malik et al., 2006; Ling et al., 2007); hence, it can be expected that NLP results may be used to annotate co-occurrence results.

CoPub Mapper used a more sophisticated information measure which took into account the distribution of entity names in the text database (Alako et al., 2005). Although Alako et al (2005) demonstrated CoPub Mapper's information measure co-relates well with microarray co-expression, the information measure was not used as a decision criterion for deciding which pairs of co-occurrences were positive results (personal communication, Guido Jenster, 2006). This is unlike 1-Mention PubGene method where all co-occurrence were taken as positive result and 5-Mention PubGene method requires at least 5 count of co-occurrence before attributing the co-occurrence as a positive result. Chen et al. (2008) used chi-square to test co-occurrence statistically to mine disease-drug interactions from clinical notes and published literature. Another possible way to calculate co-occurrence is a direct use of Poisson distribution on the assumption that co-occurrence of 2 protein names is a rare chance with respect to the entire library. Poisson distribution is a discrete distribution similar to Binomial distribution but is used for rare events, for example, to estimate the probability of accidents in a given stretch of road in a day. Poisson distribution is easier to use than Binomial distribution as it only requires the mean and does not require a standard deviation. Based on PubGene, the statistical assumption of Poisson distribution-based statistics requiring rare events (in this case, the co-occurrences of 2 protein names in a collection of text is statistically rare) can generally be held (Jenssen et al., 2001).

Although a combination of either NLP or co-occurrence in microarray analysis have been used (Li et al., 2007; Gajendran et al., 2007; Hsu et al., 2007), neither method had been used in microarray analysis for advancing lactational biology. This study attempts to examine the relation between the PubGene and Poisson distribution methods of calculating co-occurrence and explore the use of NLP-based protein-protein interaction extraction results to annotate co-occurrence results. This study also examines the use of co-occurrence analysis on 4 publically available microarray data sets on mouse lactation cycle (Master et al., 2002; Clarkson and Watson, 2003; Stein et al., 2004; Rudolph et al., 2007) as a novel hypothesis discovery tool. Master et al. (2002) used 13 microarrays to discover the presence of brown adipose tissue in mouse mammary fat pad and its role in thermoregulation, Clarkson and Watson (2003) used 24 microarrays and characterized inflammation response genes during involution, Stein et al. (2004) used 51 microarrays and discovered a set of 145 genes that are up-regulated in early involution where 49 encoded for immunoglobulins, and Rudolph et al. (2007) used 29 microarrays to study lipid synthesis in the mouse mammary gland following diets of various fat content and found that genes encoding for nutrient transporter into the cell are up-regulated following increased food intake. More importantly, each of the 4 studies independently demonstrated that the cyclical nature of mammary gland development, as observed histologically and biochemically, are reflected at the transcriptome level suggesting that microarray is a suitable tool to study the regulation of mouse lactation. It should be noted that even-though each of these microarray experiments were designed for different purposes, the principle that co-expressed genes are more functionally correlated than functionally unrelated genes remains, as demonstrated by Reverter et al. (2005).

Our results demonstrate that 5-mention PubGene method is generally statistically more significant than $99^{th}$ percentile of Poisson distribution method of calculating co-occurrence. Our results showed that 96% of the interactions extracted by NLP methods (Ling et al., 2007) overlapped with the results from 5-mention PubGene method. However, less than 2% of the microarray correlations were found in the co-occurrence graph extracted by 1-mention PubGene method. Using co-occurrence results to filter microarray co-expression correlations, we have discovered a potentially novel set of 7 protein-protein interactions that had not been previously described in the literature.

## 2. Methods

### 2.1. Microarray Datasets

The 4 microarray datasets are from Master et al. (2002) using Affymetrix Mouse Chip Mu6500 and FVB mice, Clarkson and Watson (2003) using Affymetrix U74Av2 chip and C57/BL6 mice, Rudolph et al. (2007) using Affymetrix U74Av2 chip and FVB mice, and Stein et al. (2004) using Affymetrix U74Av2 chip and Balb/C mice.

### 2.2. Co-Occurrence Calculations

Using a pre-defined list of 3653 protein names which was derived by Ling et al. (2007) from Affymetrix Mouse Chip Mu6500 microarray probeset, PubGene established 2 measures of binary co-occurrence (Jenssen et al., 2001): 1-mention method and 5 mentions method. In the 1-mention method, the appearance of 2 entity names in the same abstract will be deemed as a positive outcome whereas the 5 mentions method will require the appearance of 2 entity names in at least 5 abstracts before considered positive.

For co-occurrence modelled on Poisson distribution (Poisson co-occurrence), the number of abstracts in which both entity names appeared in is assumed to be rare as it only requires the appearance of 2 entity names within 5 articles in a collection of 10 million articles to give a precision of 0.72 (Jenssen et al., 2001). The relative occurrence frequencies of each of the 2 entities were calculated separately as a quotient of the number of abstracts in which an entity name appeared in and the total number of abstracts in the corpus. The product of relative occurrence frequency of each of the 2 entities can be taken as the mean expected probability of the 2 entities appearing in the same abstract if they are not related, which when multiplied by the total number of abstracts, can be taken as the mean number of occurrence (*lambda*) of Poisson distribution. For example, if proteinA and proteinB are found in 1000 abstracts each and there are 1 million abstracts, the relative occurrence frequency will be 0.001 each and the mean number of occurrence will be 1 ($0.001^2$ x 1000000). This means that we expect 1 abstract in a collection of 1 million to contain proteinA and proteinB if they are not related (n = 1, p = 0.5).

A positive result is where the number of abstracts in which both the 2 entities in question appeared on or above the $95^{th}$ (one-tail P < 0.05) or $99^{th}$ (one-tail P < 0.01) percentile of the Poisson distribution. In both co-occurrence calculations, entity (protein) names in text is recognized by pattern matching , as used in Ling et al. (2007).

### 2.3. Comparing Co-Occurrence and Text Processing

Two sets of comparisons were performed: within the different forms of co-occurrence, and between co-occurrence and text processing methods. The first set of comparison aims to evaluate the differences between the 3 co-occurrence methods described above. PubGene's 1-mention and 5-mentions methods were co-related singly and in combination with Poisson co-occurrence methods.

Given that the nodes (N) of a co-occurrence network represents the entities and the links or edges (E) between each node to represent a co-occurrence under the method used, the entire co-occurrence graph (G) = {N, E}, that is, a set of nodes and a set of edges. In addition, given that the same set of entities were used (same set of nodes), the differences between the 2 graphs resulted from 2 co-occurrence methods can then be simply denoted as the number of differences between the 2 sets of edges (subtraction of one set of edges with another set of edges). In practice, a total space model is used. A graph of total possible co-occurrence is where each node is "linked" or co-occurred with every node, including loops (edge to itself). Thus, a graph of total possible co-occurrence has 3653 nodes and 12694969 ($3563^2$) edges. We define a graph, G*, as the undirected graph of total possible co-occurrence without parallel edges including loops. G* has 3653 nodes and 63457030 [3563 x (3563 − 1) / 2] edges. The output graph of each co-occurrence method is reduced to the number of edges it contains as it can be assumed that the graph from 1-mention PubGene method represents the most liberal co-occurrence graph ($G_{PG1}$), the resulting graph from any other more sophisticated method ($G_i$ where i denotes the co-occurrence method) will be a proper subset of $G_{PG1}$ and certainly G*.

The second set of comparison aims at correlating co-occurrence techniques and natural language processing techniques for extracting interactions between two entities, such as two proteins. In this comparison, the extracted protein-protein binding and activation interactions, extracted using Muscorian on 860000 published abstracts using "mouse" as the keyword as previously described (Ling et al., 2007), has been used to compare against co-occurrence network of 1-Mention PubGene and 5-Mention PubGene by graph edges overlapping as described above. Briefly, Muscorian (Ling et al., 2007) normalized protein names within abstracts by converting the names into abbreviations before processing the abbreviated abstracts into a table of subject-verb-objects. Protein-protein interaction extractions were carried out by matching of each of the 12694969 ($3563^2$) pairs of protein names and verb, namely, activate or bind, in the extracted table of subject-verb-objects.

### 2.4. Mapping Co-Expression Networks onto Text-Mined Networks

A co-expression network was generated from each of the 4 in vivo data sets by pair-wise calculation of Pearson's coefficient on the intensity values across the dataset, where a coefficient of more than 0.75 or less than -0.75 signifies the presence of a co-expression between the pair of signals on the microarray (Reverter et al., 2005). The co-expression network generated from Master et al. (2002) and an intersected co-expression network generated by intersecting all 4 networks were used to map onto 1-PubGene and NLP-mined networks. For the co-expression network generated from Master et al. (2002), a 0.01 coefficient unit incremental stepwise mapping to 1-PubGene co-occurrence network as performed from 0.75 to 1 to analyze for an optimal correlation coefficient to derive a set of correlations between genes that is likely to have not been studied before (not found in 1-PubGene co-occurrence network).

## 3. Results

### 3.1. Comparing Co-Occurrence Calculation Methods

Using 3563 transcript names, there is a total of 6345703 possible pairs of interactions - 927648 (14.6%) were found using 1-Mention PubGene method and 431173 (6.80%) were found using 5-Mention PubGene method. The Poisson co-occurrence method using both $95^{th}$ or $99^{th}$ percentile threshold found 927648 co-occurrences, which is the same set as using 1-Mention PubGene method.

The mean number of co-occurrence, which is used as the mean of the Poisson distribution, is calculated as the product of the probability of occurrence of each of the entity names in the database. Using a database of 100 thousand abstracts as an example, if 500 abstracts contained the term "insulin" (500 abstracts in 100 thousand, or 0.5%) and 200 abstracts contained the term "MAP kinase" (200 abstracts in 100 thousand, or 0.2%), then the mean number of co-occurrence (lambda in Poisson distribution) is 0.001%. The range of mean number of co-occurrence for the 6345703 pairs of entities were from zero to 0.59, with mean of 0.000031. For example, if the mean is $3.1 \times 10^{-5}$, then the probability of an abstract mentioning 2 proteins not related in any functional way is $4.8 \times 10^{-10}$ or virtually zero in 6.3 million possible interactions. These results are summarized in Table 1.

|  | **Number of Clone-Pairs** | **% of Full Combination** |
|---|---|---|
| Full Combination (G*)[1] | 6345703 | 100.00 |
| 1-Mention PubGene | 927648 | 14.62 |
| 5-Mention PubGene | 431173 | 6.80 |
| Poisson Co-occurrence at 95th percentile | 927648[2] | 14.62 |
| Poisson Co-occurrence at 99th percentile | 927648[2] | 14.62 |

**Table 1 - Summary results of co-occurrence using PubGene or Poisson distribution**

[1] The undirected graph of total possible co-occurrence ($3563^2$) without parallel edges excluding self edge, which has 3653 nodes and 63457030 [3563 x (3563 – 1) / 2] edges.
[2] Same set as 1-Mention PubGene

### 3.2. Comparison of Natural Language Processing and Co-Occurrence

Natural language processing (NLP) techniques were used to extract protein-protein binding interactions and protein-protein activation interactions from almost 860000 abstracts as described in Ling et al. (2007). A total of 9803 unique binding interactions and 11365 unique activation interactions were identified, of which 2958 were both binding and activation interactions. Of the 9803 binding interactions, 9661 interactions concurred with 1-Mention PubGene method (98.55%) and 9465 interactions with 5-Mention PubGene method (96.54%). Of the 11365 activation interactions, 11280 interactions and 11111 interactions concurred with 1-Mention PubGene method (99.25%) and 5-Mention PubGene method (97.77%) respectively. Hence, of the 927648 interactions found using 1-Mention PubGene method, 1.04% (n = 9661) were binding interactions and 1.22% (n = 11280) were activation interactions. Furthermore, of the 431173 interactions found using 5-Mention PubGene method, 2.20% (n = 9465) of the interactions were binding interactions and 2.58% (n = 11111) were activation interactions. Combining binding and activation interactions (n = 18120), 1.96% of 1-Mention PubGene co-occurrence graph and 3.85% of 5-Mention PubGene co-occurrence graph were annotated respectively.

### 3.3. Mapping Co-Expression Networks onto Text-Mined Networks

Using Pearson's correlation coefficient to signify the presence of a co-expression between the pair of spots (genes) on the Master et al. (2002) data set, there are 210283 correlations between -1.00 to -0.75 and 0.75 to 1.00, of which 2014 (0.96% of correlations) are found in 1-PubGene co-occurrence network, 342 (0.16% of correlations) are found in activation network extracted by natural language processing means and 407 (0.19% of correlations) are found in binding network extracted by natural language processing means.

From incremental correlation mapping with 1-PubGene network (tabulated in Table 2 and graphed in Figure 1), there is a decline of the number of correlations from 208269 (correlation coefficient of 0.75) to 7 (correlation coefficient of 1.00). The percentage of overlap between co-occurrence and co-expression rose linearly from correlation coefficient of 0.75 to 0.85 (r = 0.959) while that of correlation coefficient of 0.86 to 0.92 is less correlated (r = 0.223). The 7 pairs of correlations in Master et al. (2002) data set with correlation coefficient of 1.00 are; lactotransferrin (Mm.282359) and solute carrier family 3 (activators of dibasic and neutral amino acid transport), member 2 (Mm.4114); B-cell translocation gene 3 (Mm.2823) and UDP-Gal:betaGlcNAc beta 1,4- galactosyltransferase, polypeptide 1 (Mm.15622); gamma-glutamyltransferase 1 (Mm.4559) and programmed cell death 4 (Mm.1605); FK506 binding protein 11 (Mm.30729) and signal recognition particle 9 (Mm.303071); FK506 binding protein 11 (Mm.30729) and Ras-related protein Rab-18 (Mm.132802); casein gamma (Mm.4908) and casein alpha (Mm.295878); G protein-coupled receptor 83 (Mm.4672) and recombination activating gene 1 activating protein 1 (Mm.17958). The amount of overlap between microarray correlations and 1-mention PubGene co-occurrence increased steadily from 0.96% at the correlation coefficient of 0.75 to 1.057% at the correlation coefficient of 0.87.

Mapping an intersect of co-expression networks of all 4 in vivo data sets (Master et al., 2002; Clarkson and Watson, 2003; Stein et al., 2004; Rudolph et al., 2007), there are 1140 correlations, of which 14 (1.23%) are found in 1-PubGene co-occurrence network, none of which corresponds to the interactions found in activation or binding networks extracted by natural language processing means (Ling et al., 2007).

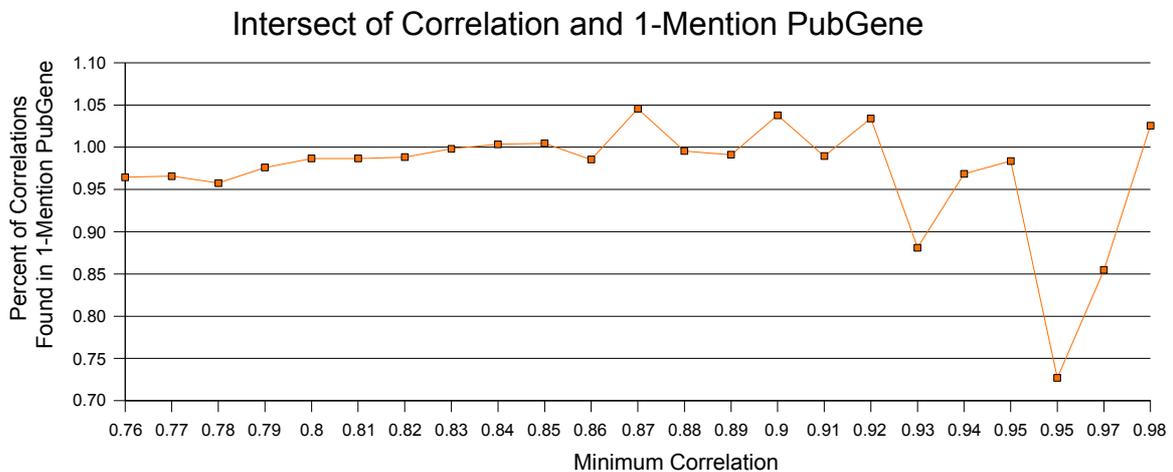

**Figure 1 – Percentage of correlation network analyzed from Maser et al. (2002) are found in 1-Mention PubGene co-occurrence**

| Minimum Correlation | Number of Correlations in Master et al. (2002) | Number of Correlations found in 1-PubGene | Percentage of Correlations Found |
|---|---|---|---|
| 0.75 | 210283 | 2014 | 0.958 |
| 0.76 | 207593 | 1983 | 0.964 |
| 0.77 | 181383 | 1735 | 0.966 |
| 0.78 | 157622 | 1495 | 0.958 |
| 0.79 | 136152 | 1316 | 0.976 |
| 0.80 | 116775 | 1141 | 0.987 |
| 0.81 | 99276 | 970 | 0.987 |
| 0.82 | 83802 | 823 | 0.988 |
| 0.83 | 70019 | 692 | 0.998 |
| 0.84 | 57872 | 575 | 1.004 |
| 0.85 | 47453 | 472 | 1.005 |
| 0.86 | 38228 | 373 | 0.985 |
| 0.87 | 30347 | 314 | 1.046 |
| 0.88 | 23740 | 234 | 0.995 |
| 0.89 | 18137 | 178 | 0.991 |
| 0.90 | 13435 | 138 | 1.038 |
| 0.91 | 9797 | 96 | 0.990 |
| 0.92 | 6849 | 70 | 1.034 |
| 0.93 | 4580 | 40 | 0.881 |
| 0.94 | 2919 | 28 | 0.969 |
| 0.95 | 1742 | 14 | 0.984 |
| 0.95 | 970 | 7 | 0.727 |
| 0.97 | 472 | 4 | 0.855 |
| 0.98 | 197 | 2 | 1.026 |
| 0.99 | 60 | 0 | 0.000 |
| 1.00 | 7 | 0 | 0.000 |

Table 2 - Summary of incremental stepwise mapping of correlation coefficients from Master et al. (2002) to 1-PubGene co-occurrence network

## 4. Discussion

Comparing the difference between PubGene (Jenssen et al., 2001) and Poisson modelling method for co-occurrence calculations, three observations could be made. Firstly, one of the common criticisms of a simple co-occurrence method as used in this study (co-occurrence of terms without considering the number of words between these terms) is that given a large number of articles or documents, every term will co-occur with every term at least once, leading to total possible co-occurrence (100% or 12694969 in this case). Our results showed that 7.31% of the total possible co-occurrence were actually found using about 860000 abstracts and only 3.40% using a more stringent method. PubGene (Jenssen et al., 2001) has also suggested that total possible co-occurrence was not evident with a much larger set of articles (10 million) and yet achieved 60% precision using only one instance of co-occurrence in 10 million articles (1-Mention PubGene) and 72% precision with 5-Mention PubGene. It can be expected with more instances of co-occurrence, precision may be higher. This might be due to the sparse distribution of entity names in the set of text as observed from the low mean number of co-occurrence used for Poisson distribution modeling. At the same time, PubGene (Jenssen et al., 2001) also illustrated that entity name recognition by simple pattern matching is able to yield quality results.

Using only results from PubGene (Jenssen et al., 2001), it can be concluded that total possible co-occurrence is unlikely for a corpus size of up to 10 million (more than half of current PubMed). Using the Poisson distribution, the mean number of co-occurrence can be expected to decrease with a larger corpus than used in this study as it is a product of the relative frequencies of each of the 2 entities. This suggests that as the size of corpus increases, it is likely that each co-occurrence of terms is more significant, suggesting that a statistical measure might be more useful in a very large corpus of more than 10 million as it takes into account both frequencies and corpus size.

Secondly, Poisson co-occurrence methods at both 95th and $99^{th}$ percentile yield the same set of results as 1-Mention PubGene method, which is expected as the maximum mean number of co-occurrence is 0.59. This implied that every co-occurrence found are essentially statistically significant in a corpus of about 860000 abstracts; thus, providing statistical basis for "1-Mention PubGene" method. This might be due to the nature of abstracts, which were known to be concise. Proteins that have no relation to each other are generally unlikely to be mentioned in the same abstract and abstracts tends to mention only crucial findings. However, the same might not apply if full text articles are used – un-related proteins could be used solely for illustrative purposes.

Thirdly, the number of co-occurrences found using 5-Mention PubGene method is substantially lower (less than half) of that by 1-Mention PubGene method which was also shown in Jenssen et al. (2001). This suggested that 5-Mention PubGene is appreciably more stringent than using Poisson co-occurrence at $99^{th}$ percentile; thus, providing statistical basis for "5-Mention PubGene" method.

Our results comparing the numbers of co-occurrence demonstrated a 50.79% decrease in co-occurrence from 1-Mention PubGene network to 5-Mention PubGene network. However, the 5-Mention PubGene network retained most of the "activation" (98.5%) and "binding" (98.0%) interactions found in 1-Mention PubGene network. This might be the consequence of 30% recall of the NLP methods (Ling et al., 2007) as it would usually require 3 or more mentions to have a reasonable chance to be identified by NLP methods. This might also be due to the observation that the 5-Mention PubGene method is more precise, in terms of accuracy, than the 1-PubGene method as shown in Jenssen et al. (2001).

The probability of a true interaction (Ling et al., 2007) existing in each of the 9661 NLP-extracted binding interactions that are also found in 1-Mention PubGene co-occurrence would be raised. The probability of a true interaction existing in each of the 9465 NLP-extracted binding interactions that are also found in 5-Mention PubGene co-occurrence would be higher. Hence, combining NLP and statistical co-occurrence techniques can improve the overall confidence of finding true interactions. However, it should be noted that statistical co-occurrence used in this work cannot raise the confidence of NLP-extracted interactions.

Nevertheless, these results also suggest that graphs of statistical co-occurrence could be annotated with information from NLP methods to indicate the nature of such interactions. In this study, 2 NLP-extracted interactions from Ling et al. (2007), "binding" and "activation", were combined. The combined "binding" and "activation" network covered 1.96% and 3.85% of 1-Mention and 5-Mention PubGene co-occurrence graph respectively. Our results demonstrate that the combined network has a higher coverage than individual "binding" or "activation" networks. Thus, it can be reasonable to expect that with more forms of interactions, such as degradation and phosphorylation, extracted with the same NLP techniques, the co-occurrence graph annotation would be more complete.

By overlapping the co-expression network analyzed from Master et al. (2002) data set to 1-Mention PubGene co-occurrence network, our results demonstrated that about 99% of the co-expression was not found in the co-occurrence network. This might suggest that the choice of Pearson's correlation coefficient threshold of more than 0.75 and less than -0.75 as suggested by Reverter et al. (2005) is likely to be sensitive in isolating functionally related genes from microarray data at the cost of reduced specificity.

Our results from incremental stepwise analysis showed that the percentage of overlap between co-expression and co-occurrence rose linearly from correlation coefficient from 0.75 to 0.85. This suggests that a correlation coefficient of 0.85 may be optimal for this data set as it is likely that using the correlation coefficient of 0.85 will result in less false positives than the correlation coefficient of 0.75. At the same time, increasing the correlation coefficient from 0.75 to 0.85 resulted in 77.4% less (47453 correlations from 210283) interaction correlations. Using this method to further describe protein-protein interactions and to generate new hypotheses, it can be argued that correlation coefficient of 0.85 will result in less false positives. While this deduction is likely as a more stringent criterion tends to reduce the rate of false positives, it is difficult to prove experimentally without exhaustive examination of each result. Nevertheless, the result suggest the possibility of using the inverse linearity of correlation coefficient and the number of gene co-expressions as a preliminary visual assessment to gauge an optimal correlation coefficient to use for a particular data set. However, on the extreme end, a correlation coefficient of 0.99 and 1.00 yielded 60 and 7 correlations respectively in Master et al. (2002) data set but none was found in 1-Mention PubGene co-occurrence network. This suggests that high-throughput genomic techniques such as microarrays, present a vast amount of un-mined biological information that had not been examined experimentally.

By exploring the literature for the biological significance for each of the 7 pairs of perfectly co-expressed genes using Swanson's method (Swanson, 1990), it was found that all 7 pairs were biologically significant. Lactotransferrin (Ishii et al., 2007) and solute carrier family 3 (activators of dibasic and neutral amino acid transport), member 2 (Feral et al., 2005) were involved in cell adhesion. B-cell translocation gene 3 (Guehenneux et al., 1997) and UDP-Gal:betaGlcNAc beta 1,4-galactosyltransferase, polypeptide 1 (Mori et al., 2004) were involved in cell cycle control. Casein gamma and casein alpha are well-established components of milk. Gamma-glutamyltransferase 1 (Huseby et al., 2003) and programmed cell death 4 (Frankel et al., 2008) were known to be regulating apoptotic pathways. Rab18 (Vazquez-Martinez et al., 2007), signal recognition particle 9 (Egea et al., 2004) and FK506 binding protein 11 (Dybkaer et al., 2007) were known to be involved in the secretory pathway. G protein-coupled receptor 83 (Lu et al., 2007) and recombination activating gene 1 activating protein 1 (Igarashi et al., 2001) were known to be involved in T-cell function. Taken together, these suggest that the set of 7 correlations have not likely been described and may prove to be valuable new hypotheses in the study of mouse mammary physiology. It is also plausible that this argument can be extended to the set of 53 highly co-expressed genes (0.99 < correlation coefficient < 1.00).

Intersecting the 4 in vivo data sets into a co-expression network increases the power of the analysis as it represents correlation among gene expression that are more than 0.75 or less than -0.75 in all 4 data sets. There were 1140 examples of co-expression in this intersect and only 14 co-expressions (1.23%) were found in the one-mention PubGene co-occurrence network, but none in either the binding or activation networks extracted by natural language processing. This suggests that these 14 co-expressions are neither binding nor activating interactions. Textpresso (Muller et al., 2004) had defined a total of 36 molecular associations between 2 proteins which includes binding and activation. Future work will expand NLP mining to 34 other interactions to improve the annotation of co-occurrence networks.

Reverter et al. (2005) had previously analysed 5 microarray data sets by expression correlation and demonstrated that genes of related functions exhibit similar expression profile across different experimental conditions. Our results suggest 1126 co-expressed genes across 4 microarray data sets are not found in the co-occurrence network. This may be a new set of valuable information in the study of mouse mammary physiology as these pairs of genes have not been previously mentioned in the same publication and experimental examination of these potential interactions is needed to understand the biological significance of these co-expressions.

## 5. Conclusions

We conclude that the 5-mention PubGene method is more stringent than the $99^{th}$ percentile of Poisson distribution method. In this study, we demonstrate the use of a liberal co-occurrence-based literature analysis (1-Mention PubGene method) to represent the state of research knowledge in functional protein-protein interactions as a sieve to isolate potentially novel hypotheses from microarray co-expression analyses for further research.

## Authors' contributions

ML, CL and KRN contribute equally to the design of experiments and analysis of results. ML carries out the experiments.

**Appendix A – Use of Python in this work**

Python programming had been used throughout this study, which had been incorporated into Muscorian (Ling et al., 2007). The following are code snippets to demonstrate the calculation of Poisson distribution and the intersection of Master et al., 2002 and 1-mention PubGene results as shown in Figure 1 and Table 2.

Given that *muscopedia.dbcursor* is the database cursor and *pmc_abstract* table to contain the abstracts, the Poisson distribution model for each pair of entity (gene or protein) names is constructed by the function *commandJobCloneOccurrencePoisson*,

```
class Poisson:
    mean = 0.0
    def __init__(self, lamb = 0.0): self.mean = lamb

    def factorial(self, m):
        value=1
        if m != 0:
            while m !=1:
                value=value*m
                m=m-1
        return value

    def PDF(self, x):
        return math.exp(self.mean)* \
            pow(self.mean,x)/self.factorial(x)

    def inverseCDF(self, prob):
        cprob = 0.0
        x = 0
        while (cprob < prob):
            cprob = cprob + self.PDF(x)
            x = x + 1
        return (x, cprob)

def commandJobCloneOccurrencePoisson(self):
    poisson = Poisson()
    muscopedia.dbcursor.execute('\
      select count(pmid) from pmc_abstract')
    abstractcount = \
      float(self.muscopedia.dbcursor.fetchall()[0][0])
    muscopedia.dbcursor.execute('\
      select jclone, occurrence from jclone_occurrence')
    dataset = [[clone[0].strip(), clone[1]] for clone in
               self.muscopedia.dbcursor.fetchall()]
    muscopedia.dbcursor.execute("\
      delete from jclone_occur_stat")
    count = 0
    for subj in dataset:
        for obj in dataset:
            mean = (float(subj[1])/abstractcount)* \
                    (float(obj[1])/abstractcount)
            poisson.mean = mean
            (poi95, prob) = poisson.inverseCDF(0.95)
            (poi99, prob) = poisson.inverseCDF(0.99)
            count = count + 1
            sqlstmt = "insert into jclone_occur_stat (clone1,\
                clone2, randomoccur, poisson95, poisson99) \
                values ('%s','%s','%.6f','%s','%s')" % \
```

```
                            (str(subj[0]), str(obj[0]), mean, \
                            str(poi95), str(poi99))
                try: muscopedia.dbcursor.execute(sqlstmt)
                except IOError: pass
                if (count % 1000) == 0:
                    muscopedia.dbconnect.commit()
```

Each pair of entities was searched in each abstract using SQL statements, such as "*select count(pmid) from pmc_abstract where text containing 'insulin' and 'MAPK'*", and the number of abstracts found was matched against *jclone_occur_stat* table for statistical significance based on the calculated Poisson distribution.

The results were exported from muscopedia (Muscorian's database) as a tab-delimited file and analyzed using the following code to generate Table 2:

```
import sets

lc = open('lc_cor.csv','r').readlines()
lc = [x[:-1] for x in lc]
lc = [x.split('\t') for x in lc]
d = {}
for x in lc:
    try: t = d[(x[1], x[0])]
    except KeyError: d[(x[0], x[1])] = float(x[2])

lc = [(x[0], x[1], d[x]) for x in d]
l = [(x[0], x[1]) for x in d]
l = sets.Set(l)

def process_sif(file):
    a = open(file,'r').readlines()
    a = [x[:-1] for x in a]
    a = [x.split('\tpp\t') for x in a]
    return [(x[0], x[1]) for x in a]

a = sets.Set(process_sif('pubgene1.sif'))

print "# intersect of pubgene1.sif and LC data: " + \
    str(len(l.intersection(a)))
print "# LC data not in pubgene1.sif: " + \
    str(len(l.difference(a)))
print "# pubgene1.sif not in LC data: " + \
    str(len(a.difference(l)))
print ""

cor = 0.74
while (cor < 1.0):
    t = [(x[0], x[1]) for x in lc if x[2] > cor]
    l = sets.Set(t)
    cor = cor + 0.01
    print "LC correlation: " + str(cor)
    print "# intersect of pubgene1.sif and LC data: " + \
        str(len(l.intersection(a)))
    print "# LC data not in pubgene1.sif: " + \
        str(len(l.difference(a)))
    print "# pubgene1.sif not in LC data: " + \
        str(len(a.difference(l)))
    print ""
```

**Appendix B – PubGene algorithm and its main results**

PubGene (Jenssen et al., 2001) algorithm is a count-based algorithm which simply counts the number of abstracts with both entity names. Using "insulin" and "MAPK" as the pair of entities, PubGene algorithm can be implemented using the following SQL, "*select count(pmid), 'insulin', 'MAPK' from pmc_abstract where text containing 'insulin' and text containing 'MAPK'*". 1-Mention PubGene and 5-Mention PubGene can be isolated by filtering for *count(pmid)* to be more than zero and four respectively. PubGene (Jenssen et al., 2001) had demonstrated that the precision of 1-Mention is 60% while the precision of 5-Mention is 72%.